# INFORMATIONAL FIELD
# AND SUPERLUMINAL COMMUNICATION

## V.P. Oleinik

Department of General and Theoretical Physics,
National Technical University of Ukraine "Kiev Polytechnic Institute",
Prospect Pobedy 37, Kiev, 03056, Ukraine; e-mail: yuri@arepjev.relc.com
http://www.chronos.msu.ru/lab-kaf/Oleynik/eoleynik.html

*Abstract.* The field of scalar and vector potentials in electrodynamics is shown to represent an informational field capable of superluminally transmitting a signal (information) with no energy and momentum transfer. This conclusion strictly follows from Maxwell's equations for electromagnetic field interacting with electric charges and currents in vacuum, without resort to any additional hypotheses. That superluminal communication is possible is seen from the fact that the own field, generated by particles and inseparable from them, transforms the environment into a special physical medium which is capable to instantaneously transfer a signal (information) about any changes, happening to a particle in the region of its basic localization, to arbitrarily large distances. The phenomenon of superluminal communication is caused by the non-local connection of scalar and vector potentials with the electric and magnetic field strengths. The basis for the mechanism of superluminal communication considered in this work is the Aharonov-Bohm effect indicative of the field of electromagnetic potentials as a real physical field, which directly influences the behaviour of electron waves. The conclusion is made that in quantum systems superluminal signals occur constantly, in any quantum processes. The occurrence of superluminal signals is due to the space-time symmetry breaking of a special kind, consisting in that the equations for potentials do not possess relativistic invariance though Maxwell's equations for the field strengths are Lorentz-invariant. The results presented do not contradict the physical principles underlying special relativity and confirm the fundamental conclusion, made for the first time by de Broglie, that gauge invariance is not an absolute law in physics.

## 1. Introduction

According to the generally accepted point of view, the velocity of light in vacuum is the greatest possible velocity of transfer of a signal existing in nature. This conclusion was formulated by A.Einstein as a consequence of the special theory of relativity (STR) and of the causality principle as follows: "... There is no way of sending the signals which would propagate faster than light in vacuum" (see [1], p. 157).

Meanwhile, research of relativistic wave equations shows that those admit solutions which describe the waves propagating with the group velocity exceeding the velocity of light in vacuum [2-10]. It should be emphasized that the occurrence of superluminal signals does not result in violation of causality principle. As appears from the detailed analysis [2,11,12], the standard arguments that the existence of superluminal signals is forbidden by the causality principle are erroneous because those are based exclusively on the use of Lorentz transformations, i.e. of kinematics. The cause and effect relationship between events is caused, however, by the process of interaction transfer from one event to another and, hence, is a problem of dynamics. This process can be described merely with the help of solutions of dynamical equations obeying certain initial conditions. It is essential that because of the difference in the initial conditions corresponding to the process of propagating superluminal signal in various inertial frames of reference there is no way to transfer information to the past [2]. As shown in [12], the conclusion that superluminal signals cannot exist does not follow from STR and is an additional hypothesis inconsistent with Maxwell's equations.

The prediction that electromagnetic waves exist whose group velocity exceeds the velocity of light in vacuum is supported by numerous experimental results [13-19]. Besides, mention should be made of the paper [20] in which experimental data are given on the superluminal information transfer in artificially created media (metamaterials) with negative dielectric and magnetic susceptibilities, and also the review papers [21-23] in which various aspects of the problem of superluminal movements are considered.

On the basis of theoretical research on the superluminal communication problem, one can conclude that there are no basic restrictions that would prohibit him from producing the pulses of electromagnetic field with group velocity exceeding that of light in vacuum. On the other hand, it is evident from experimental research that by now **the light barrier has been surmounted** and the signal velocity achievable is limited only by the potentialities of experimental equipment.

The experimental works mentioned above are dealt with superluminal propagation of optical pulses, i.e. the packets of transverse electromagnetic waves. Mathematically, such packets correspond to solutions of homogeneous wave equations. Study of the non-homogeneous wave equations describing electromagnetic field shows that the field consists of two components – the photon component and the own field of electrically charged particles, which essentially differ from each other by their physical characteristics [12,24-26]. The own field is inseparable from the charged particles generating it; it is a field of standing waves of matter, which is not subject to the wave-corpuscular duality and cannot be reduced to a stream of photons. This field transforms the environment into a physical medium, which is capable to instantaneously transfer a signal (information) about a perturbation occurring at some point of space to arbitrarily large distance. The conclusion that superluminal communication with the help of own field is possible follows from the formulation of quantum electrodynamics [12,27,28] in which the self-action of electron is taken into account – the back influence on electron of the own field generated by it. Owing to the self-action, electron turns into an open self-organizing system, which, in view of the long-range character of the own field, occupies the whole space. For such a system to be stable, there should exist a physical mechanism binding its parts into a single whole. It is the instantaneous transfer of information via the own field that is, apparently, such a mechanism.

The conclusion that the instantaneous long-range interaction and the short-range interaction occurring with the velocity of light can co-exist simultaneously is made also in papers [29-31], starting from the completely different line of reasoning - on the basis of the separated-potential method in classical electrodynamics (the electrodynamics dualism concept) [29,30] and on the basis of the relativistic quantum field theory [31].

The existence of superluminal signals is supported by the astronomical observations, which have been conducted for the first time by N.A. Kozyrev [32,33]. According to Kozyrev's investigation, there exists in nature an action-at-a-distance of one body on another resulting in the superluminal transfer of signal. These results have been confirmed by M.M. Lavrentjev, I.A. Eganova and others [34,35] and also repeated in part in [36]. In the papers mentioned above the Kozyrev conclusions have been specified and have received further generalization and development.

The field of scalar and vector potentials is shown **i**n the present paper to be **an informational field** which is capable to transfer information about physical process with the speed lying in the range from zero to infinity, the information transfer being not accompanied, generally speaking, by the energy and momentum transfer from the generator of signal to the receiver [37].

The physical mechanism of superluminal communication is due to the non-local connection of scalar and vector potentials with the electric field strength $E$ and magnetic induction $B$. As in quantum mechanics the interaction of electromagnetic field with charged particles is described in terms of potentials rather than of electric $E$ and magnetic $B$ fields, in view of the non-locality mentioned above, the change of fields $E$ and $B$ at the moment of time $t$ in some bounded region of space $\Gamma$ can lead to the change of potentials at the later moment $t + 0$ at an observation point separated from the region $\Gamma$ by any distance. Owing to the change of potentials, there occurs a shift of the wave function phases of charged particles at the observation point, which can be detected by the shift of the interference fringes.

The basis for the mechanism of superluminal communication being considered is the Aharonov-Bohm effect [38] indicative of the field of magnetic potential is a real physical field, which manifests itself in the experiments on electron diffraction in the presence of magnetic field. As is emphasized by G. Lochak in [39], "this remarkable effect, which proves the influence of a fieldless magnetic potential on electron waves, is shocking for those who have been convinced for a century that electromagnetic potentials are only mathematical intermediate entities". The shift of the interference fringes in the Aharonov-Bohm effect is caused by the change in the wave function phase of electron in magnetic field, which follows directly from the de Broglie formula linking the wavelength of electron with its velocity and vector potential. It is essential that the electron interference depends on the choice of the vector potential gauge, the dependence of interference fringes on potential manifesting itself even in the case that the source of potential is outside the region sandwiched between the electronic trajectories resulting in interference [39].

The main result of the present paper is the proof that the existence of superluminal signals, whose carrier is the own field of electrically charged particles, strictly follows from Maxwell's equations for the electromagnetic field interacting with electric charges and currents in vacuum. In a sense, **superluminal signals and their physical carrier as a special physical medium are derived from the equations of electromagnetic field.**

Superluminal signaling with the help of the own field of charged particles represents a qualitatively new means of communication possessing a lot of essential advantages over the optical one. It should be emphasized that at present all the necessary prerequisites are available for the creation of the means of superluminal communication based on the use of own fields [40].

The results obtained in the paper are the following.

In section 2 a family of potentials of electromagnetic field is received which describes the transfer of information in space at any speed. The characteristic is given of the basic physical properties of the own field of potentials as a carrier of superluminal signals. The own field of electrically charged particles is shown to turn the surrounding space into a special physical medium which "is so arranged" that it is capable to instantaneously transfer a signal (information) about any changes occurring to a particle in the region of its basic localization to arbitrarily large distance.

In section 3 an insight is given into why it follows necessarily from Maxwell's equations that superluminal signals exist and also the physical mechanism of occurrence of these signals is considered. The prospects for the further research are pointed out in the field of quantum electrodynamics taking properly into account the own fields and the superluminal signals associated with them.

The paper represents a development of physical ideas and generalization of the results set forth in the monograph [12].

## 2. The Field of Potentials as an Informational Field[1]

---

[1] In what follows we use the system of units in which $c = 1$ ($c$ is the velocity of light in vacuum).

We proceed from the set of Maxwell's equations for the electric $\boldsymbol{E}$ and magnetic $\boldsymbol{B}$ fields generated in vacuum by electric charges and currents:

$$\left[\vec{\nabla}\boldsymbol{E}\right]=-\frac{\partial \boldsymbol{B}}{\partial t}, \quad \vec{\nabla}\boldsymbol{B}=0, \quad \left[\vec{\nabla}\boldsymbol{B}\right]=\frac{\partial \boldsymbol{E}}{\partial t}+4\pi\boldsymbol{j}, \quad \vec{\nabla}\boldsymbol{E}=4\pi\rho, \tag{1}$$

with $\rho$ and $\boldsymbol{j}$ being the charge and current densities. The solution to equations (1) can be looked for, as usual, in the form

$$\boldsymbol{E}=-\frac{\partial \boldsymbol{A}}{\partial t}-\vec{\nabla}\varphi, \quad \boldsymbol{B}=\left[\vec{\nabla}\boldsymbol{A}\right], \tag{2}$$

where $\boldsymbol{A}$ and $\varphi$ are the vector and scalar potentials. We use the following constraint ($\varepsilon$ is a real parameter):

$$\varepsilon\frac{\partial \varphi}{\partial t}+\vec{\nabla}\boldsymbol{A}=0. \tag{3}$$

Equality (3) coincides at $\varepsilon=0$ with the Coulomb gauge and at $\varepsilon=1$ with the Lorentz gauge. Because of this, the constraint (3) can be termed the generalized gauge.

When substituting the relationships (2) into equations (1), the first two of the equations become identities, and the second two, in view of the generalized gauge, can be represented in the form

$$\left(\varepsilon\frac{\partial^2}{\partial t^2}-\vec{\nabla}^2\right)\boldsymbol{A}=\left[\vec{\nabla}\boldsymbol{B}\right]-\varepsilon\frac{\partial \boldsymbol{E}}{\partial t}, \tag{4}$$

$$\left(\varepsilon\frac{\partial^2}{\partial t^2}-\vec{\nabla}^2\right)\varphi=4\pi\rho, \tag{5}$$

where the quantities $\left[\vec{\nabla}\boldsymbol{B}\right]$ and $\frac{\partial \boldsymbol{E}}{\partial t}$ are supposed to be related to each other by the third of the equations (1).

By considering equalities (4) and (5) as the equations for potentials, solution to these equations can be written in terms of the Green function $G^{(\varepsilon)}(x-x')$:

$$\boldsymbol{A}(x)=\int d^4 x'\, G^{(\varepsilon)}(x-x')\left(\left[\vec{\nabla}'\boldsymbol{B}(x')\right]-\varepsilon\frac{\partial}{\partial t'}\boldsymbol{E}(x')\right)\equiv \boldsymbol{A}^{(\varepsilon)}(x),$$
$$\varphi(x)=4\pi\int d^4 x'\, G^{(\varepsilon)}(x-x')\rho(x')\equiv \varphi^{(\varepsilon)}(x), \tag{6}$$

where Green's function obeys the equation

$$\left(\varepsilon\frac{\partial^2}{\partial t^2}-\vec{\nabla}^2\right)G^{(\varepsilon)}(x-x')=\delta^4(x-x'). \tag{7}$$

In the above has been used the notation: $\boldsymbol{A}(x)=\boldsymbol{A}(\boldsymbol{r},t)$, $\varphi(x)=\varphi(\boldsymbol{r},t)$, $d^4 x=dt\, d\boldsymbol{r}$, $x=(t,\boldsymbol{r})$ is the four-dimensional radius-vector, $\delta^4(x-x')=\delta(t-t')\delta(\boldsymbol{r}-\boldsymbol{r}')$ is the four-dimensional Dirac $\delta$-function. Further we shall assume, for definiteness, that $\varepsilon\geq0$ and that $G^{(\varepsilon)}(x-x')$ is the retarded Green function. In this case Green's function is given by

$$G^{(\varepsilon)}(x-x')=\frac{1}{4\pi|\boldsymbol{r}-\boldsymbol{r}'|}\delta\left(t-t'-\sqrt{\varepsilon}|\boldsymbol{r}-\boldsymbol{r}'|\right). \tag{8}$$

Let the fields $\boldsymbol{E}$ and $\boldsymbol{B}$ be distinct from zero only in some bounded region of the 4-space $\Gamma$. Then, using the last of Maxwell's equations (1) and expression (8) for Green's function, we can write equalities (6) as

$$A^{(\varepsilon)}(x) = \frac{1}{4\pi}\int\limits_{\Gamma} d^4x' \frac{1}{|\boldsymbol{r}-\boldsymbol{r}'|}\delta\left(t-t'-\sqrt{\varepsilon}\,|\boldsymbol{r}-\boldsymbol{r}'|\right)\left(\vec{\nabla}'\boldsymbol{E}(x'),\left[\vec{\nabla}'\boldsymbol{B}(x')\right]-\varepsilon\frac{\partial}{\partial t'}\boldsymbol{E}(x')\right), \qquad (9)$$

with $A^{(\varepsilon)}(x) = \left(\varphi^{(\varepsilon)}(x), \boldsymbol{A}^{(\varepsilon)}(x)\right)$. Formula (9) relates the potentials of electromagnetic field to the fields $\boldsymbol{E}$ and $\boldsymbol{B}$, and thus is a generalization of the known formula of electrostatics

$$\varphi(\boldsymbol{r}_1) - \varphi(\boldsymbol{r}_2) = \int\limits_{r_1}^{r_2}\boldsymbol{E}(\boldsymbol{r})\,d\boldsymbol{r}$$

to the arbitrary electromagnetic field. As can be seen from (2) and (9), the fields $\boldsymbol{E}$ and $\boldsymbol{B}$ are expressed in terms of potentials $\boldsymbol{A}$ and $\varphi$ locally, and the inverse relation of the potentials to the fields $\boldsymbol{E}$ and $\boldsymbol{B}$ is non-local both in space and in time: to define the potentials at a point $\boldsymbol{r}$ at an instant of time $t$ it is necessary to know the behaviour of electric and magnetic fields in the whole space at the preceding moments of time.

If the region $\Gamma$ reduces to the vicinity of some point $x_0 = (t_0, \boldsymbol{r}_0)$, then, according to (9), the information obtained from electric $\boldsymbol{E}$ and magnetic $\boldsymbol{B}$ fields at the point $x_0$ is transferred by the field of potentials to an observation point $x$ at the velocity $1/\sqrt{\varepsilon}$ (in usual units $c/\sqrt{\varepsilon}$, $c$ is the velocity of light in vacuum). In this case, generally speaking, at the observation point $x$ the fields $\boldsymbol{E}$ and $\boldsymbol{B}$ can be absent, though $A^{(\varepsilon)}(x) \neq 0$. If, for example, $\varphi^{(\varepsilon)}(x) = f(t) \neq 0$, $\boldsymbol{A}^{(\varepsilon)}(x) = \vec{\nabla}g(\boldsymbol{r}) \neq 0$, with $f(t)$ and $g(\boldsymbol{r})$ being some scalar functions, then, according to formulas (2), $\boldsymbol{E}(x) = \boldsymbol{B}(x) = 0$, but $A^{(\varepsilon)}(x) \neq 0$. From this it follows that the field of potentials $A^{(\varepsilon)}(x)$ can transfer information about physical process from one point of 4-space to the other with no transfer of the energy, momentum and other physical quantities determined by the fields $\boldsymbol{E}$ and $\boldsymbol{B}$. Thus, **the field $A^{(\varepsilon)}(x)$ plays the role of an informational field**, the velocity of information transfer exceeding that of light in vacuum at $\varepsilon < 1$.

Expression (9) for scalar and vector potentials does not allow the velocity to be derived, at which energy and momentum are transferred in electromagnetic field, as the substitution of potentials (9) into (2) results in the identities of the type: $\boldsymbol{E} = \boldsymbol{E}, \boldsymbol{B} = \boldsymbol{B}$. To establish the magnitude of this velocity, split equation (4) into two independent equations, one of which describes vortex ($\boldsymbol{A}_\perp$) and the other potential ($\boldsymbol{A}_\parallel$) components of vector-potential. The decomposition of any vector field $\boldsymbol{a} = \boldsymbol{a}(\boldsymbol{r}, t)$ into potential and vortex components ($\boldsymbol{a} = \boldsymbol{a}_\perp + \boldsymbol{a}_\parallel$) can be performed by the following formulas:

$$\boldsymbol{a}_\parallel(\boldsymbol{r}, t) = -\frac{1}{4\pi}grad\,div\int d\boldsymbol{r}'\,\frac{1}{|\boldsymbol{r}-\boldsymbol{r}'|}\boldsymbol{a}(\boldsymbol{r}', t),$$

$$\boldsymbol{a}_\perp(\boldsymbol{r}, t) = \frac{1}{4\pi}curl\,curl\int d\boldsymbol{r}'\,\frac{1}{|\boldsymbol{r}-\boldsymbol{r}'|}\boldsymbol{a}(\boldsymbol{r}', t).$$

$$(10)$$

Taking into account, that according to Maxwell's equations (1) $4\pi\boldsymbol{j}_\parallel = -\dfrac{\partial\boldsymbol{E}_\parallel}{\partial t}$ and, in view of (2),

$\boldsymbol{E}_\perp = -\dfrac{\partial\boldsymbol{A}_\perp}{\partial t}$, one can easily find equality (4) to be equivalent to the following two equations:

$$\left(\frac{\partial^2}{\partial t^2}-\vec{\nabla}^2\right)\boldsymbol{A}_\perp = 4\pi\,\boldsymbol{j}_\perp, \quad \left(\varepsilon\frac{\partial^2}{\partial t^2}-\vec{\nabla}^2\right)\boldsymbol{A}_\parallel = 4\pi\varepsilon\,\boldsymbol{j}_\parallel. \qquad (11)$$

Solution of equations (11) can be represented as:

$$\boldsymbol{A}_\perp(x) = 4\pi\int d^4x'\,G^{(1)}(x-x')\boldsymbol{j}_\perp(x'), \qquad (12)$$

$$A_{\parallel}(x) = 4\pi\varepsilon \int d^4x'\, G^{(\varepsilon)}(x-x')\boldsymbol{j}_{\parallel}(x') \equiv A_{\parallel}^{(\varepsilon)}(x). \tag{13}$$

Alternatively, the relationships (12) and (13) can be obtained by splitting vector-potential $A^{(\varepsilon)}(x)$ (6) into vortex and potential components with the use of formulas (10). As can be seen from simple calculation, $A_{\perp}^{(\varepsilon)}(x) = A_{\perp}(x)$, i.e. the quantity $A_{\perp}^{(\varepsilon)}(x)$ is independent of parameter $\varepsilon$. Introducing the notation $A_{\parallel}^{(\varepsilon)}(x) = \left(\varphi^{(\varepsilon)}(x), A_{\parallel}^{(\varepsilon)}(x)\right)$, one can arrive at the following expression (see the second of formulas (6)):

$$A_{\parallel}^{(\varepsilon)}(x) = 4\pi \int d^4x'\, G^{(\varepsilon)}(x-x')\left(\rho(x'),\, \varepsilon\, \boldsymbol{j}_{\parallel}(x')\right). \tag{14}$$

Expressions (12) and (14) allow one to calculate the fields $\boldsymbol{E}$ and $\boldsymbol{B}$ by formulas (2) and to establish thereby the dependence of the fields and their components on the 4-current density $j(x) = \left(\rho(x), \boldsymbol{j}(x)\right)$. From formulas (2), (12) and (14) it follows that if the functions $\rho(x), \boldsymbol{j}_{\parallel}(x)$ and $\boldsymbol{j}_{\perp}(x)$ were concentrated at a point, the magnetic field $\boldsymbol{B}$ and the vortex component $\boldsymbol{E}_{\perp}(x)$ of electric field would transfer information with the velocity of light in vacuum and potential component $\boldsymbol{E}_{\parallel}(x)$ of electric field with the velocity $1/\sqrt{\varepsilon}$.

Calculate the quantities $\boldsymbol{j}_{\parallel}(x)$ and $\boldsymbol{j}_{\perp}(x)$ for a point-like particle of charge $q$ moving in a path $\boldsymbol{r} = \boldsymbol{r}_0(t)$. In this case

$$j(x) = q(1, \boldsymbol{v}_0(t))\, \delta(\boldsymbol{r} - \boldsymbol{r}_0(t)), \tag{15}$$

where $\boldsymbol{v}_0(t) = \dfrac{d\boldsymbol{r}_0(t)}{dt}$ is the velocity of the particle. Potential component of the current density can be calculated with the use of the first of formulas (10) and of the continuity equation $\partial\rho/\partial t + \vec{\nabla}\boldsymbol{j} = 0$. Simple calculations result in the following generic formula:

$$\boldsymbol{j}_{/\!/}(\boldsymbol{r}, t) = \frac{1}{4\pi}\vec{\nabla}\frac{\partial}{\partial t}\int d\boldsymbol{r}'\, \frac{1}{|\boldsymbol{r} - \boldsymbol{r}'|}\rho(\boldsymbol{r}', t). \tag{16}$$

Using (15), one can obtain for a point-like particle the following formula:

$$\boldsymbol{j}_{/\!/}(\boldsymbol{r}, t) = \frac{q}{4\pi}\vec{\nabla}\frac{\partial}{\partial t}\frac{1}{|\boldsymbol{r} - \boldsymbol{r}_0(t)|} \equiv \boldsymbol{j}_{/\!/}(x). \tag{17}$$

The vortex component of current can be defined as

$$\boldsymbol{j}_{\perp}(x) = \boldsymbol{j}(x) - \boldsymbol{j}_{/\!/}(x). \tag{18}$$

As is seen from (15), (17) and (18), although the quantity $\boldsymbol{j}(x)$ in the case of a point-like particle is localized at some point of space, its potential and vortex components are distributed over the whole space. This means that it is impossible to attribute a finite velocity of propagation in space to the physical fields corresponding to components $A_{\parallel}^{(\varepsilon)}(x)$ and $A_{\perp}(x)$ of the vector-potential (see (12) and (13)). Because of the fact that the current density components are distributed over all space, the latter **is instantaneously endowed with information** about any physical process in which the point-like particle participates.

It can be shown that potentials $A^{(\varepsilon)}(x)$ and $A^{(\varepsilon')}(x)$, where $A^{(\varepsilon)}(x) = \left(\varphi^{(\varepsilon)}(x), \boldsymbol{A}^{(\varepsilon)}(x)\right)$, at $\varepsilon \neq \varepsilon'$ are related to each other by gauge transformation

$$A^{(\varepsilon)}(x) = A^{(\varepsilon')}(x) + \left(\partial/\partial t, -\vec{\nabla}\right)g(x), \tag{19}$$

with $g(x)$ being some continuous function. From here it follows that the potentials above are in fact physically equivalent in the sense that they result in the same fields $\boldsymbol{E}$ and $\boldsymbol{B}$.

Taking into account formulas (12) and (13), calculate the total vector-potential of electromagnetic field:

$$A^{(\varepsilon)}(x) = 4\pi \int d^4x' \left( G^{(1)}(x-x') \boldsymbol{j}_\perp(x') + G^{(\varepsilon)}(x-x') \varepsilon \boldsymbol{j}_{\|}(x') \right). \tag{20}$$

In view of (14) and (20), for the Lorentz gauge it can be derived the following representation for potential:

$$A^{(1)}(x) = 4\pi \int d^4x' G^{(1)}(x-x') j(x'). \tag{21}$$

As is obvious from (15) and (21), in the case of point-like particle the field of potentials in Lorentz's gauge and the field strengths $\boldsymbol{E}$ and $\boldsymbol{B}$ are propagated in space with the same velocity – the velocity of light. Further, relying on the equivalence (in the mentioned above sense) of potentials $A^{(\varepsilon)}(x)$ and $A^{(\varepsilon')}(x)$ at $\varepsilon \neq \varepsilon'$, one can conclude that for point-like particle the field strengths are always propagated with the velocity of light irrespective of the choice of gauge. However, the velocity of propagation of the field of potentials, i.e. the velocity of transfer of physical information not accompanied by energy and momentum transport in space, essentially depends on gauge and can vary within the interval $(0, \infty)$.

Thus, in the electromagnetic field generated by classical point-like particles, information is transferred with the velocity of light only in the event that potentials obey the Lorentz gauge. If parameter $\varepsilon$ is plotted on a coordinate axis, the solutions of Maxwell's equations subject to the Lorentz gauge correspond merely to one point. Obviously, such solutions make up only a very narrow class of solutions of Maxwell's equations. The solutions corresponding to generalized gauge at $\varepsilon < 1$ describe such electromagnetic fields, which are capable to transfer information with superluminal velocity.

The statement made above that solutions of Maxwell's equations, corresponding to the Lorentz gauge, describe the electromagnetic field propagating with the velocity of light is obviously of an approximate character: it is true only in the view of standard formulation of quantum electrodynamics [41] based on the concept of point-like particles. The formulation, as is evident from an analysis of the problem [12,27], is a rather rough approximation to the reality because it takes no account of the self-action of particles – the back action of the own field generated by a particle on the same particle. The self-action of particles is not small in comparison with interaction between particles and, consequently, it cannot be taken into account by perturbation theory. Owing to the incorporation of self-action into theoretical scheme the particles cease to be point-like and become spatially extended objects. As a result the potentials in the Lorentz gauge, too, describe superluminal transfer of information. In other words, in describing real electromagnetic interaction **there is no gauge of potentials which would allow one "to get rid" of superluminal signals.**

To elucidate the physical nature of the occurrence of superluminal signals, we address again to Maxwell's equations (1). Splitting each of the vectors involved in equations (1) into potential and vortex components according to formulas (10), we present these equations as two independent subsets: for vortex electric $\boldsymbol{E}_\perp$ and magnetic $\boldsymbol{B}$ fields -

$$\left[ \vec{\nabla} \boldsymbol{E}_\perp \right] = -\frac{\partial \boldsymbol{B}}{\partial t}, \quad \vec{\nabla} \boldsymbol{B} = 0, \quad \left[ \vec{\nabla} \boldsymbol{B} \right] - \frac{\partial \boldsymbol{E}_\perp}{\partial t} = 4\pi \boldsymbol{j}_\perp, \quad \vec{\nabla} \boldsymbol{E}_\perp = 0, \tag{22}$$

and for potential electric field $\boldsymbol{E}_{\|}$ -

$$\frac{\partial}{\partial t} \boldsymbol{E}_{\|} = -4\pi \boldsymbol{j}_{\|}, \quad \vec{\nabla} \boldsymbol{E}_{\|} = 4\pi \rho, \quad \left[ \vec{\nabla} \boldsymbol{E}_{\|} \right] = 0. \tag{23}$$

According to (22), the electromagnetic field $\left( \boldsymbol{E}_\perp, \boldsymbol{B} \right)$ does not contain sources and sinks and is generated by vortex current $\boldsymbol{j}_\perp$. The latter, being specified at some space-time point, generates at this point a whirlwind of magnetic field and a time-dependent electric field ( $\left[ \vec{\nabla} \boldsymbol{B} \right] - \frac{\partial \boldsymbol{E}_\perp}{\partial t} \neq 0$ ), the occurrence of a whirlwind of electric field at some place being accompanied, as can be seen from the first of the equations (22), by the change of magnetic field with time at the same place. Because the electromagnetic field $(\boldsymbol{E}_{\perp}, \boldsymbol{B})$ is of vortex character, the subset (22) at $\boldsymbol{j}_{\perp} = 0$ has non-trivial solutions corresponding to free electromagnetic waves.

According to (23), the charge density $\rho = \rho(\boldsymbol{r}, t)$ is a source (or a sink) of potential field $\boldsymbol{E}_{\|}$ and defines its dependence on time. At $\rho = 0$ (in this case, too, $\boldsymbol{j}_{\|} = 0$, see equality (16)) the subset (23) has merely a trivial solution ($\boldsymbol{E}_{\|}(\boldsymbol{r}, t) = const$) not having physical sense. From the physical point of view, this means that potential field $\boldsymbol{E}_{\|}$ is generated by electrically charged particles, is intimately connected with them, and cannot exist in their absence. The field $\boldsymbol{E}_{\|}$ is, thus, not an independent degree of freedom of electromagnetic field.

From the above analysis of Maxwell's equations for the potential and vortex components of electromagnetic field, it can be inferred that **the electromagnetic field** generated by electric charges and currents in vacuum **consists of two components**: 1) **the own field** of electrically charged particles which is not an independent degree of freedom of electromagnetic field, and 2) **the field of transverse electromagnetic waves**. The potential component of the own field is described by potential $A_{\|}^{(c)}(x)$ (14), and the vortex one is described, in part, by potential $A_{\perp}(x)$ (12). In addition, the field $A_{\perp}(x)$ contains the transverse electromagnetic waves, which are emitted by and separated from particles when they move with acceleration (the radiation field [42]). These components essentially differ from each other by their physical characteristics.

The own field is, in a sense, a constituent of charged particles and does not submit to the wave-corpuscular duality. It represents a field of standing waves of matter rigidly connected to particles and going from one particle to another or to infinity. This field is purely of a classical character and does not reduce to the set of photons. The own field of a particle seems to be responsible for the occurrence of its wave properties, which manifest themselves, for example, in the experiments on electron diffraction.

In distinction to the own field, the field of transverse electromagnetic waves represents a stream of photons moving in vacuum with the velocity of light. It is that part of the electromagnetic field generated by charged particles, which is split out of the own field and then separated from particles.

To gain a better insight into the relationship between the photon component of electromagnetic field and the own field, imagine a point-like particle at rest in an inertial frame of reference. The own field of such a particle reduces to static Coulomb field that is potential. When the particle moves uniformly and rectilinearly, a vortex component is split out of the own field in the whole space [12]. This component represents, obviously, such a vortex electromagnetic field, which cannot be a stream of photons. At last, when the particle moves with acceleration, its vortex own field is distorted, with the result that a stream of elementary excitations of the field is generated which turn to quanta of electromagnetic field at large distances from the particle.

The existence of the own field seems to be a necessary condition for the field of radiation as a stream of photons to occur. In fact, the own field, which is indissolubly related to particles and inseparable from them, serves as a physical medium in which photons are formed and travel and without which they cannot exist.

### 3. The Physical Mechanism of Superluminal Communication

In the previous section, based on the analysis of Maxwell's equations, the family of potentials of the electromagnetic field generated in vacuum by electric charges and currents is derived. The potentials depend on a real parameter $\varepsilon$, which is involved both in the D'Alembert operator $\varepsilon \, \partial^2 / \partial t^2 - \bar{\nabla}^2$, describing wave properties of the field, and in the constraint imposed on the potentials. The distinguishing feature of potentials is that in all the gauges being considered, except for the Lorentz gauge, both the components of vector-potential, potential and vortex, taken separately, and the total vector-potential depend on superluminal signals.

The constraint (3), imposed on the components of potential $A_i^{(\varepsilon)}$ ($i = 0,1,2,3$), at $\varepsilon \neq 1$ is not invariant under Lorentz transformations and, consequently, the quantity $A^{(\varepsilon)}(x) \equiv \left( A_0^{(\varepsilon)}(x), \boldsymbol{A}^{(\varepsilon)}(x) \right)$ at $\varepsilon \neq 1$ is not a 4-vector. Despite this, the quantity $F_{\mu\nu}^{(\varepsilon)}(x) = \partial_\mu A_\nu^{(\varepsilon)}(x) - \partial_\nu A_\mu^{(\varepsilon)}(x)$, where $\partial_\mu = \partial / \partial x^\mu$, $x^\mu = (t, \boldsymbol{r})$, represents a 4-tensor (the 4-tensor of electromagnetic field), and in consequence the relativistic invariance of Maxwell's equations (1) is retained. **The occurrence of superluminal signals can be seen from the above to be associated with the space-time symmetry breaking of a special sort**, consisting in that the equations for potentials do not possess relativistic invariance though Maxwell's equations for the field strengths (1) are Lorentz-invariant. The above solutions of Maxwell's equations describing superluminal transfer of information are characterized, thus, by the breaking of space-time symmetry of the field of potentials.

With the help of gauge transformation (19) it is not difficult to separate out that part of potential that is responsible for the symmetry breaking:

$$A^{(\varepsilon)}(x) = A^{(1)}(x) + \left( \partial / \partial t, -\bar{\nabla} \right) g(x) \, . \tag{24}$$

Here $A^{(1)}(x)$ is the potential in the Lorentz gauge that represents a 4-vector, and the second addend in the right-hand side of (24) is not a 4-vector and takes into account the symmetry breaking of the field of potentials. Obviously, the potential (24) would be a 4-vector if the quantity $g(x)$ were a scalar, however this quantity, as is seen from an analysis [37], is not a scalar.

By virtue of the fact that the constraint (3) at $\varepsilon \neq 1$ is not invariant under Lorentz transformations, the question arises of whether the solutions of the equations of electromagnetic field with broken symmetry have physical sense. Perhaps, such solutions should be discarded as they break the ban imposed by special relativity on superluminal signals? Notice, first of all, that the Coulomb gauge ($\varepsilon = 0$) represents a special case of symmetry breaking of the equations for potentials. However, it is generally believed that the use of the Coulomb gauge is quite possible [43]. As Maxwell's equations completely govern the behaviour of electromagnetic field interacting with charged particles, it is natural to believe that all their solutions subject to required conditions (solutions should be continuous together with a certain number of their partial derivatives, bounded and uniquely determined - these conditions are called standard) have physical sense. It is not difficult to check that the solutions to the equations for potentials in the generalized gauge satisfy the standard conditions at any real magnitudes of $\varepsilon$. From here it follows that there are no grounds to discard them and, hence, one can hope that all these solutions are realized in nature.

In the Lorentz gauge the electromagnetic field is described by the 4-potential $A^{(1)}(x)$, which will be denoted by $A(x) = \left( A_0(x), \boldsymbol{A}(x) \right)$. Its components satisfy equations

$$\left( \frac{\partial^2}{\partial t^2} - \bar{\nabla}^2 \right) A_i(x) = 4\pi \bar{j}_i(x) \quad (i = 0,1,2,3) \tag{25}$$

and constraint

$$\frac{\partial A_0(x)}{\partial t} + \bar{\nabla} \boldsymbol{A}(x) = 0 \, . \tag{26}$$

At the first sight, from the expressions (25) and (26), whose advantages are simplicity, symmetry with respect to the components of potential, and explicit relativistic invariance, follows of necessity the conclusion that electromagnetic field propagates in vacuum with the velocity of light $c$. Really, equalities (25) are the wave equations in which the D'Alembert operator points to the velocity of light as the velocity of propagation of electromagnetic waves and the right part, in the case of point-like charged particles, is described by the functions concentrated at separate geometrical points. Under these conditions, apparently, the velocity of electromagnetic signals can be equal solely to that of light.

However this conclusion is erroneous. To explain why the occurrence of superluminal signals in electromagnetic field is inevitable, we shall decompose electromagnetic field into potential and vortex components and write the equations (25) in the form corresponding to such a decomposition:

$$\left(\frac{\partial^2}{\partial t^2} - \vec{\nabla}^2\right) A_0(x) = 4\pi\rho(x), \quad \left(\frac{\partial^2}{\partial t^2} - \vec{\nabla}^2\right) \boldsymbol{A}_\|(x) = 4\pi \boldsymbol{j}_\|(x),$$

$$\left(\frac{\partial^2}{\partial t^2} - \vec{\nabla}^2\right) \boldsymbol{A}_\perp(x) = 4\pi \boldsymbol{j}_\perp(x). \tag{27}$$

Here the first two equations describe the potential component of the field, and the last describes the vortex one.

As is emphasized in the previous section, the potential electric field $\boldsymbol{E}_\|$ is not an independent degree of freedom of electromagnetic field: the field $\boldsymbol{E}_\|$, described in the Lorentz gauge by the components $A_0$ and $\boldsymbol{A}_{\|}$, is generated by electrically charged particles and is inseparable from them. For this reason in the consistent quantum theory this field should be included in the definition of particle at the very initial stage of constructing the theory as it is done in [12,27,28]. In such a formulation of the quantum theory, with due allowance for the self-action of particles, the electric charge of particle (for example, of electron) proves to be distributed over all space: the distribution of electron charge consists of the region of the basic localization and of a tail extending from this region to infinity. As is seen from the results of this paper and of the papers [12,25,26], the "smearing" of the particle's charge in space automatically gives rise to the occurrence of superluminal signals.

Thus, the inaccuracy of the conclusions, which are usually drawn on the basis of equations (25) and (26), is connected with the fact that these conclusions prove to be purely formal, divorced from physical reality. They do not take into account the fact that potential component of electromagnetic field, inextricably related to electric charge of particles, represents, in essence, an integral part of electrically charged particles and, therefore, it is intolerable to consider it as an independent degree of freedom of electromagnetic field.

The Lorentz gauge is the only gauge, which does not give rise to symmetry breaking of the equations for potentials and thereby masks the existence of superluminal signals. Simplicity, symmetry, and explicit relativistic invariance of the equations for potentials in this gauge seem to have merely a formal character, not reflecting the true physical nature of electromagnetic field.

One can readily see that in any gauge the quantity $\left(A_0^{(\varepsilon)}, A_{//}^{(\varepsilon)}\right)$ describing potential component of the own field depends on superluminal signals. If the self-action of particle is properly allowed for, the potential component of the own field is expressed in terms of the particle's wave function [12]. From this it follows immediately that the dynamic equation describing self-action of particle takes of necessity into account the occurrence of superluminal signals in the particle's field. Obviously, the mechanism of the superluminal signal transfer consists in the non-local dependence of the dynamic equation on time and spatial coordinates.

Note that the potentials $A^{(\varepsilon)}$ at different values of $\varepsilon$, connected with each other by gauge transformations, are physically equivalent only in the sense that they correspond to the same strengths of electromagnetic field. As in classical mechanics the interaction between the field and particles is described in terms of the field strengths, it is natural to tell about equivalence from the point of view of classical dynamics. In quantum mechanics, however, the interaction is described in terms of potentials and not in terms of strengths. For this reason, from the point of view of quantum dynamics, the potentials $A^{(\varepsilon)}$, corresponding to different values of $\varepsilon$, are not equivalent to each other. They differently describe the interaction of microparticles with electromagnetic field, as it is seen from electron interference in magnetic field, and result in various velocities of information transfer. The conclusion made for the first time by de Broglie that gauge invariance is not an absolute law in physics (see [39]) is, thus, supported not only by the Aharonov-Bohm effect, but also by the dependence of the velocity of information transfer on the choice of gauge of potentials.

It should be emphasized that superluminal signals must of necessity occur in electromagnetic field interacting with charged quantum particles. Really, in quantum electrodynamics the 4-current density is expressed by (see [41])

$$j(x) = e\overline{\Psi}(x)(\gamma_0, \vec{\gamma})\Psi(x), \tag{28}$$

where $e$ is the electric charge of electron, $\Psi(x)$ is the electron field operator, $\overline{\Psi} = \Psi^+\gamma_0$, $\gamma_0$ and $\vec{\gamma}$ are Dirac's $\gamma$-matrixes. In view of the fact that the field operator $\Psi(x)$ is not localized at a geometrical point but distributed in some way over all space, the electromagnetic field, the interaction of which with electron field is governed by Maxwell's equations (1) with the 4-current density (28), is, obviously, the carrier of superluminal signals.

Let us now turn to the physical mechanism of the occurrence of superluminal signals. First of all, recall once again that in quantum mechanics the interaction of electromagnetic field with charged particles is described not in terms of the fields $\boldsymbol{E}$ and $\boldsymbol{B}$ but in terms of the scalar and vector potentials [41]. This interaction results in the shift of phases of the wave functions of particles, which causes the shift of interference fringes in electron diffraction experiment. The phase displacements of wave functions contain all physical information about interacting systems, which can be transferred to any distance without the energy and momentum transfer.

Consider an electric charge $q$, which is at rest at the moment of time $t$ at a point $\boldsymbol{r} = \boldsymbol{r}_0(t)$ in some inertial reference frame. If within the interval of time $(t, t_1)$, with $t_1 > t$, the charge is displaced from this point, in some vicinity of it there will occur a magnetic field $\boldsymbol{B}$ and a time-dependent electric field $\boldsymbol{E}$. According to formula (9), as applied to the limiting case $\varepsilon \to +0$, a trial charge $q'$ placed at an observation point $\boldsymbol{r}$ is subjected to the electromagnetic field described by the potential

$$A^{(0)}(\boldsymbol{r}, t) = \frac{1}{4\pi}\int d\boldsymbol{r}' \frac{1}{|\boldsymbol{r} - \boldsymbol{r}'|}\left(\overline{\nabla}'\boldsymbol{E}(\boldsymbol{r}', t), \left[\overline{\nabla}'\boldsymbol{B}(\boldsymbol{r}', t)\right]\right). \tag{29}$$

As is seen from (29), the change of the fields $\boldsymbol{E}$ and $\boldsymbol{B}$ in the vicinity of point $\boldsymbol{r}_0(t)$, caused by the displacement of the charge $q$, will lead to the change in potentials at point $\boldsymbol{r}$, which can be situated at any distance from $\boldsymbol{r}_0(t)$. As a result, at the moment of time $t_1 + 0$ the phase of wave function of the particle placed in a vicinity of the observation point $\boldsymbol{r}$ will change. This means that the trial charge $q'$ "will feel" the displacement of the charge $q$ even in the event that the points $\boldsymbol{r}_0(t)$ and $\boldsymbol{r}$ are separated by a space-like interval, i.e. under the conditions when the light signal caused by acceleration of the charge $q$ has no time to reach the observation point and, hence, the phase displacement of wave function takes place in the absence of the energy and momentum transfer from the location point of the charge $q$ to the observation point. Superluminal transfer of information is caused, thus, by the non-local relation of the potentials to the fields $\boldsymbol{E}$ and $\boldsymbol{B}$.

Results of this paper testify that superluminal signals inevitably occur in describing the interaction of electromagnetic field with charged particles on the basis of Maxwell's equations. This raises the question of how it was possible not to notice superluminal signals over almost hundred years elapsed after creating STR? The answer is that

1. superluminal signals were, in a sense, ideologically forbidden since the time of creating STR as a subject contradicting the physical principles and, therefore, this topic was not worked out properly;

2. the requirement that the equations for potentials be relativistically invariant reliably masks superluminal signals: as may be inferred from an analysis of the problem, if the Lorentz gauge is used and, in addition, the electric charges are treated as point-like particles and calculations are carried out within the standard perturbation theory (see [41]), the contributions to the 4-potential from superluminal excitations cancel out mutually in any order of perturbation theory, so that no evidence of superluminal signals can be found.

It is necessary to emphasize that in quantum systems, because the 4-current density is described in them by the expressions of the type (28), superluminal signaling occurs constantly, in any quantum processes. Physically, this is due to the fact that the carrier of superluminal signals is the own field of potentials which forms a special physical medium filling the whole space and capable of transferring information instantaneously to any distances from one material object to another.

The own field of particle seems to contain four components according to the four kinds of interactions known at present - electromagnetic, weak, strong, and gravitational. Each of these components is a classical field connecting the particle with surroundings with the help of superluminal signals. The role of the own field of potentials in organization of the world is that it transforms particles and bodies into the open self-organizing systems whose stability is provided by the interaction with environment with the help of superluminal signals. Note that the own field as a physical medium has little in common with physical vacuum of standard quantum field theory (see [41]). One of essential differences is that the own field has purely classical character while physical vacuum "is occupied" by virtual quantum particles - photons, electrons, electron-positron pairs and others.

As Descartes believed, there is nothing in the world except for the ether and its whirlwinds. According to results of this paper, investigation of Maxwell's equations for electromagnetic field allow one to proceed from philosophical reasoning on the ether as a special physical medium to the detailed studying of physical properties and behaviour of the medium at the strict mathematical level. Almost in all investigations, which were carried out in the twentieth century in the field of quantum electrodynamics, the Lorentz gauge was used, which allowed one to disguise superluminal signals. Studying electromagnetic field with the use of generalized gauge at $\varepsilon \neq 1$ will open up a new, mysterious world in which the key role is played by superluminal signals. Solution of the dynamic equations taking properly into account the own fields will allow one to establish the physical mechanisms governing quantum processes with allowance for superluminal signals, to predict the new physical effects connected with these mechanisms, and thereby to point the new ways to the development of engineering.

The results obtained are indicative of the possibility of creating a qualitatively new means of communication that is based on the use of superluminal signals carried by the own fields of charged particles. By their physical characteristics – by the speed and the range of information transfer, by the capacity to penetrate through obstacles, by their reliability in service - the new communication facilities, working on superluminal signals, will be much superior to the now existing ones.

The author is grateful to Arepjev Yu.D., Abakarov D.I., Ovcharuk M.E. and Tretjak O.V. for interest in the paper, useful remarks and stimulating discussions.